%% file: samplepaper.tex
\documentclass[runningheads]{llncs}
\usepackage[T1]{fontenc}
\usepackage{graphicx}
\usepackage{hyperref}
\usepackage{orcidlink}

\begin{document}
\newcommand{\IDchun}{0009-0002-1516-7586}
\newcommand{\IDzhuang}{0000-0002-0089-2454}
\newcommand{\IDsutedjo}{0000-0001-8737-5769}
\newcommand{\IDxu}{0009-0008-9419-3988}
\newcommand{\IDren}{0009-0006-2967-3438}
\newcommand{\IDxia}{0000-0002-2676-9032}

\title{ArguMath: AI-Simulated Environment for Pre-Service Teacher Training in Orchestrating Classroom Mathematics Argumentation}

\institute{Texas A\&M University, College Station, TX 77843, USA
\email{\{jiwonchun,ylzhuang,asutedjo,xucol000,rren,mengxia\}@tamu.edu}}

\author{Jiwon Chun\orcidlink{\IDchun} \and
Yuling Zhuang\orcidlink{\IDzhuang} \and
Armanto Sutedjo\orcidlink{\IDsutedjo} \and
Colin Xu\orcidlink{\IDxu} \and
Rong Ren\orcidlink{\IDren} \and
Meng Xia\orcidlink{\IDxia}}

\authorrunning{J. Chun et al.}

%
\titlerunning{ArguMath: AI-Simulated Environment for Pre-Service Teacher Training}

\maketitle

\begin{abstract}
Facilitating productive mathematical argumentation, especially asking rational questions, is essential yet remains challenging for pre-service mathematics teachers (PMTs), who often have limited opportunities to apply abstract theoretical knowledge in authentic practice. At the same time, recent advances in large language models (LLMs) have expanded the potential for simulating students in educational settings, enabling low-risk environments for instructional practice. To inform the design of a system that supports PMTs in orchestrating classroom argumentation, we conducted a formative study with eight experienced mathematics teachers to identify key design requirements, including personalization, realistic simulations, structured reflection, and ease of use.
Building on these requirements, we developed \textit{ArguMath}, an AI-simulated classroom environment that supports PMTs in practicing the orchestration of mathematical argumentation. \textit{ArguMath} comprises three core components: (1) customization of classroom settings; (2) simulation of classroom discussions with AI-based students grounded in authentic transcripts and augmented with real-time instructional suggestions; and (3) structured reflection through discourse annotation and overall feedback. Results from an exploratory user study with seven PMTs, complemented by interviews with four experienced teachers, indicate that \textit{ArguMath} has the potential to support PMTs' classroom orchestration skills, particularly theory-aligned questioning strategies.

\end{abstract}
\input{sections/1-intro}
\input{sections/3-formative}
\input{sections/4-system}
\input{sections/5-evaluation}
\input{sections/6-results}
\input{sections/7-discussion}
\input{sections/8-conclusion}
%

%
%
%
\bibliographystyle{splncs04}
\bibliography{sn-bibliography}
\end{document}

%% file: sections/1-intro.tex
\section{Introduction}

Mathematical argumentation, a process in which students construct and justify claims with evidence, is fundamental to enhancing mathematical understanding and problem-solving skills~\cite{conner2014teacher}. To foster students' productive engagement, teachers must effectively orchestrate discourse by posing purposeful questions that elicit and connect student reasoning~\cite{brown2017using}. However, pre-service mathematics teachers (PMTs) often lack sufficient opportunities to practice these complex facilitation skills due to limited class time and the high demands of individualized instructor feedback~\cite{wagner2014using}. Current preparation relies heavily on passive observation, which is often insufficient to bridge the gap between theoretical knowledge and classroom enactment~\cite{arseven2018use,wess2025pre}. Therefore, there is a pressing need for low-risk, authentic training environments where PMTs can actively rehearse instructional strategies to prepare for the complexities of real classroom contexts~\cite{wagner2013using}.

LLMs have transformed classroom simulations, offering risk-free environments for teaching innovation~\cite{xu2023leveraging,jin2025teachtune}. While applications facilitate engaging teacher training, they primarily simulate small student groups, often resulting in idealized or misaligned interactions with the nuances of real classroom practice~\cite{pan2025tutorup,xu2025classroom}.
Furthermore, few studies focus on the orchestration of mathematical argumentation as a core instructional practice for PMTs. Prior research has largely emphasized general performance feedback rather than integrating established pedagogical frameworks into real-time guidance and reflection~\cite{wess2025pre}.

To address these gaps, we introduce \textit{ArguMath}, an AI-based classroom simulator designed to scaffold PMTs' argumentation orchestration skills. Informed by a formative study with eight in-service mathematics teachers on the practice of argumentation for PMTs, \textit{ArguMath} has three components: (1) a personalized interface to set parameters including grade level, topic, and student knowledge; (2) authentic argumentation simulations powered by Retrieval-Augmented Generation (RAG) using TIMSS Video\footnote{\href{https://www.timssvideo.com/}{https://www.timssvideo.com/}}; and (3) pedagogical integration of frameworks (Toulmin's argumentation model~\cite{toulmin2003uses}, Teacher Rational Questioning Framework (TRQF)~\cite{zhuang2022teachers}) to provide real-time guidance and structured reflection.

To evaluate \textit{ArguMath}, we compared it against a baseline system that reflects traditional PMT preparation, in which participants review scenarios and receive generic, non-interactive feedback. An exploratory user study with seven PMTs and four experienced teachers was conducted to address two research questions. \textbf{RQ1}: How does PMTs' argumentation orchestration, specifically questioning performance, differ between \textit{ArguMath} and the baseline? \textbf{RQ2}: How do PMTs and experienced teachers perceive \textit{ArguMath} compared to the baseline system? Results have suggested evidence that \textit{ArguMath} better supports PMTs' classroom orchestration and theory-aligned questioning strategies. Participants also rated the simulation and reflection features as highly useful and intuitive, emphasizing their value for professional development.

%% file: sections/3-formative.tex
\section{Formative Study and Design Requirements}
To inform the system design for PMTs' argumentation orchestration, we interviewed eight experienced mathematics teachers familiar with PMT training (6 female, 2 male) with an average of $18.13$ years of experience ($SD = 11.29$), covering Grades 6--12 and class sizes of 4--32. Each one-hour Zoom interview covered five areas: current classroom discourse practices, novice teacher challenges, PMT support needs, AI experiences, and feedback on a early-stage prototype which is designed a chat interface with multiple AI-simulated students. Participants reviewed the interface and provided suggestions. 

Based on our formative study, we distilled four key design requirements. \textbf{DR1.} Provide a personalized interface to adjust the simulation environment for training, including grade level, math topic, and classroom dynamics. Teachers emphasized that adapting strategies to diverse classroom compositions is a core challenge for PMTs. \textbf{DR2.} Deliver realistic AI-based simulations to help PMTs navigate unpredictable classroom dynamics. Participants valued rehearsing with students of varied understanding and engagement levels, as well as requested immediate feedback on questioning clarity and probing depth. \textbf{DR3.} Integrate structured and theory-based reflection activities guided by pedagogical frameworks to analyze discourse patterns. Participants suggested that annotating teacher-student interactions helps PMTs identify and refine their orchestration strategies. \textbf{DR4.} Ensure an intuitive user experience with visual cues (e.g., emojis) to facilitate quick perception of student engagement. Reducing cognitive burden allows PMTs to focus on the complexity of argumentation.

%% file: sections/4-system.tex
\section{System}
Integrating the design requirements, we present \textit{ArguMath} to support PMTs in practicing classroom argumentation facilitation. Leveraging OpenAI's GPT-4o, the system employs prompt engineering to achieve a three-step workflow.

\textbf{\textit{Step~1: Context Personalization}}
PMTs first configure the classroom simulation by specifying three parameters: \textit{Grade Level}, the simulated students' grade; \textit{Math Topic}, the instructional content; and \textit{Class Description}, for which PMTs provide a brief summary of student demographics, engagement, and prior mathematical understanding (e.g., ``about half of students are highly engaged, while others struggle with applying algebraic concepts''). These inputs generate a personalized pedagogical context that governs student behaviors and interaction logic in subsequent steps.

\begin{figure}
    \centering
    \includegraphics[width=0.96\linewidth]{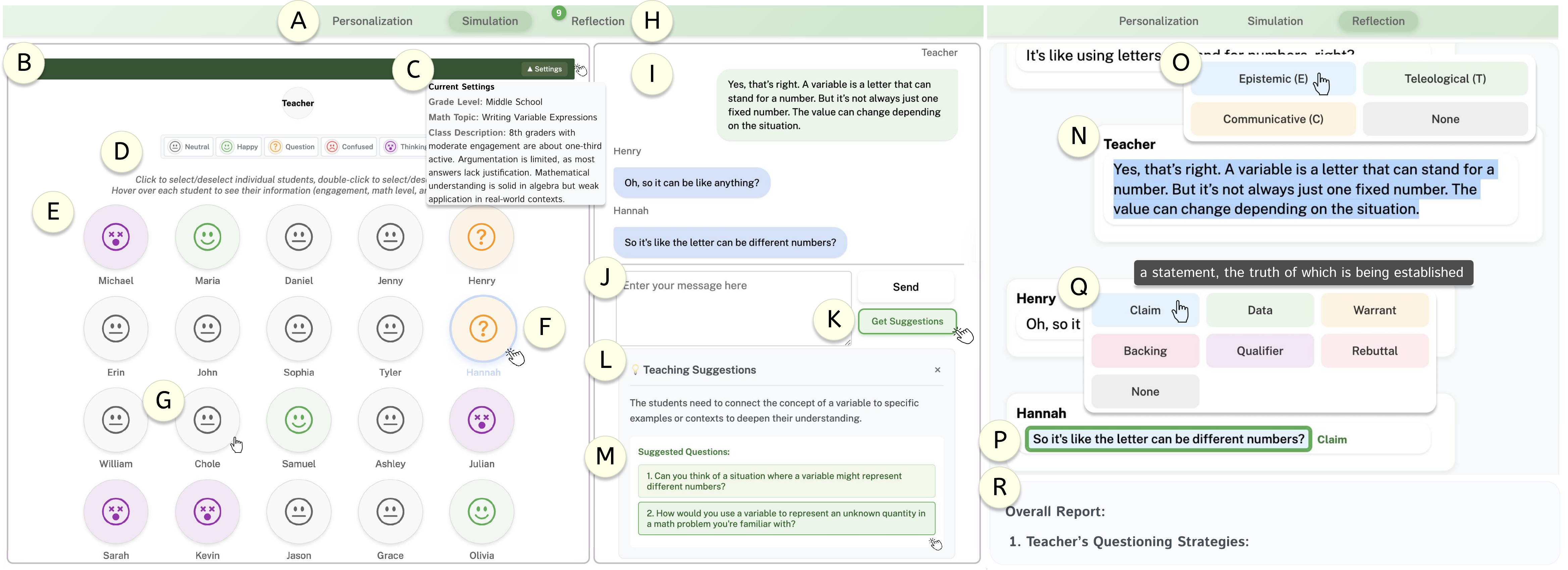}
    \caption{Interface of \textit{Step~2: Classroom Simulation} (left) and \textit{Step~3: Strategy Reflection} (right). (A) Navigation bar. (B) Virtual classroom with AI-simulated students. (C) \textit{Settings} from \textit{Step~1}. (D) Emoji and interaction legends. (E) Student faces and speech bubbles. (F) Student selection controls. (G) Hover view of student profiles. (H) Responding student count. (I) Full chat view. (J) Text input. (K) \textit{Get Suggestions} for questioning strategies feedback. (L) Teaching strategy suggestions. (M) Recommended questions. (N) Full chat from \textit{Step~2}. (O, Q) Dialogue labeling using TRQF and Toulmin categories. (P) Label confirmation. (R) Overall feedback.}
    \label{fig:step2}
    \end{figure}

\textbf{\textit{Step~2: Classroom Simulation}}
Reflecting the average class size from our formative study ($M=24$), we simulated a classroom of twenty students (Figure~\ref{fig:step2}). While initial prompt-based generation resulted in formal and unnatural responses (e.g., ``It's a convention in algebra''), we transitioned to Retrieval-Augmented Generation (RAG), leveraging authentic classroom transcripts~\cite{lewis2020retrieval}. \textit{ArguMath} generates more natural utterances characterized by human-like hesitation and explanatory reasoning (e.g., ``We don't include the multiplication sign because it can look confusing next to a variable, like the letter x''). 

    \textit{Dataset Construction} To generate a realistic simulation, we utilized publicly available classroom videos and transcripts from TIMSS Video\footnote{\href{https://www.timssvideo.com/}{https://www.timssvideo.com/}}. Twelve eighth-grade mathematics lessons with English transcripts were selected. 
    Two authors segmented the transcripts and annotated them with speaker roles. We then constructed 214 unique student profiles by using an LLM to extract and summarize each student's participation patterns (Teacher call, Voluntary, Mixed), engagement level (Low, Medium, High), mathematical level (Beginner, Beginner-Intermediate, Intermediate, Intermediate-Proficient, Proficient), argumentation level (None, Statement only, Simple reasoning, Partial reasoning, Justification, Application reasoning, Reasoning with justification, Guidance reasoning, Clarification), and typical utterances from the transcripts~\cite{wu2025embracing}. These profiles were subsequently verified and refined by the two authors.

    \textit{Student Selection} The system leverages RAG to select 20 student profiles from a 214 dataset based on \textit{Step~1} settings. Selection criteria include engagement, mathematical proficiency, and argumentation levels. At onset, these profiles are mapped to virtual identities (Figure~\ref{fig:step2}, left).
    
    \textit{Student Response and Emoji Generation} At the start of the simulation, students remain silent with neutral emojis. Once a teacher poses a question (Figure~\ref{fig:step2} J), \textit{ArguMath} generates real-time student responses and affective emojis (Figure~\ref{fig:step2} E, I). If the teacher does not specify respondents, the system selects them based on the established classroom context and student profiles. To ensure authenticity, generated utterances incorporate fillers, contractions, and expressions of confusion or excitement. In parallel, \textit{ArguMath} updates student emojis (neutral, happy, curious, confused, thinking). This allows PMTs to quickly perceive classroom participation.

    \textit{Suggestion} \textit{ArguMath} provides real-time guidance on questioning strategies by integrating the classroom context and conversation history (Figure~\ref{fig:step2} L, M). The system evaluates teacher questions using the TRQF and analyzes student responses using Toulmin's model. Integrating these pedagogical frameworks and classroom context, \textit{ArguMath} generates a single-sentence reasoning suggestion. To maintain clarity, the system omits explicit framework references while providing two recommended questions for the next turn.


\textbf{\textit{Step~3: Strategy Reflection}}
After the simulation, PMTs annotate the dialogue to analyze their facilitation strategies. They then access overall feedback and improvement suggestions (Figure~\ref{fig:step2}, right).

    \textit{Annotation and Overall Feedback} Using an integrated interface~\cite{chun2025conflictlens}, PMTs label teacher questions based on the TRQF (\textit{Epistemic}, \textit{Teleological}, \textit{Communicative}) and student responses using Toulmin's elements (\textit{Claim}, \textit{Data}, \textit{Warrant}, \textit{Backing}, \textit{Qualifier}, \textit{Rebuttal}) (Figure~\ref{fig:step2} O, Q). The system provides immediate verification of these labels; a domain expert evaluation of two 20-turn dialogues confirmed an annotation accuracy of approximately 95\%. After PMTs submit a brief self-reflection, the system provides suggestions for improvement and supports follow-up inquiries (Figure~\ref{fig:step2} R).

%% file: sections/5-evaluation.tex
\section{Evaluation}
To address RQ1 and RQ2, we conducted an exploratory user study with seven PMTs and interviews with four experienced mathematics teachers. 

\subsection{User Study}
\textit{\textbf{Participants}}
We recruited seven female PMTs ($M_{age}=21$, U1--U7) from an undergraduate teacher education program. Participants reported AI usage ($3$ often, $2$ sometimes, $2$ rarely), general comfort with new technologies ($5/7$ positive). Their teaching experience averaged 0.64 years ($SD=0.48$).

\textit{\textbf{Baseline}}
We created a baseline system which follows standard PMT training by focusing on the analysis of classroom transcripts. It provides theoretical guidance on the TRQF and Toulmin's model, followed by text-only transcripts from the TIMSS Video (e.g., \textit{Writing Variable Expressions} or \textit{Ratios and Division}). Participants analyze these transcripts using the provided frameworks through an input field, with correct answers displayed only after submission.

\textit{\textbf{Procedure}} Following a within-subjects design, participants completed two training cycles using the baseline and \textit{ArguMath} in counterbalanced order~\cite{pan2025tutorup}. Each session was followed by post-task surveys; while the baseline survey focused on usability, the \textit{ArguMath} survey included step-specific evaluations. All testing sessions utilized \textit{ArguMath} with the \textit{Suggestion} feature disabled to evaluate independent facilitation skills. 
The study was controlled at the middle school level (Grades 6--8) with assigned topics (e.g., \textit{Writing Variable Expressions}, \textit{Ratios and Division}) and class descriptions. Training contexts featured moderate engagement and strong algebraic foundations. In contrast, testing contexts featured lower engagement with diverse mathematical proficiency levels. The simulation starts with a math problem to solve, and participants were instructed to prioritize facilitating mathematical argumentation over task completion.

\textit{\textbf{Testing Performance Evaluation}}
 To compare system effectiveness, we analyzed testing dialogues and student prompting frequencies for all participants (U1--U7) across both conditions. Two researchers independently coded teacher questions and student responses. Discrepancies were resolved through iterative discussion until a consensus was reached.

\subsection{Experienced Teacher Interview}
We conducted interviews with experienced mathematics teachers to evaluate \textit{ArguMath}'s workflow, feature utility, and potential for pedagogical improvement. Four experienced female mathematics teachers were recruited and their teaching experience averaged 12.25 years ($SD = 4.03$), covering middle (Grades 6--7) and high school (Grades 9--12) with class sizes of 4--30 students. While three experts (E1--E3) participated in the formative study, E4 was newly recruited. Following a system introduction, experts explored \textit{ArguMath} using self-selected parameters for grade level, math topic, and class description in \textit{Step~1}. After the session, participants completed a questionnaire evaluating the overall workflow, specific features, and potential areas for future improvement. 

%% file: sections/6-results.tex
\subsection{Results}
To address RQ1 and RQ2, we report quantitative analyses of interaction logs and post-task ratings, followed by qualitative findings from expert interviews.

\begin{table*}[ht]
\small
\centering
\caption{Performance of seven participants (U1--U7) comparing the baseline system (left) and \textit{ArguMath} (right) in terms of elicited student responses, teacher questions (TRQF categories), and student responses (Toulmin elements).}
\scriptsize
\resizebox{\textwidth}{!}{
\begin{tabular}{|c|c|ccc|cccccc||c|ccc|cccccc|}
\hline
& \multicolumn{10}{c||}{\textbf{Baseline}} & \multicolumn{10}{c|}{\textbf{\textit{ArguMath}}} \\
\hline
\textbf{Id} & \textbf{N}
            & \multicolumn{3}{c|}{\textbf{TRQF}}
            & \multicolumn{6}{c||}{\textbf{Toulmin}}
            & \textbf{N}
            & \multicolumn{3}{c|}{\textbf{TRQF}}
            & \multicolumn{6}{c|}{\textbf{Toulmin}} \\
&  & E & T & C & Cl & Da & Wa & Ba & Qu & Re
&  & E & T & C & Cl & Da & Wa & Ba & Qu & Re \\
\hline
U1 & 4 & 2 & - & 2 & 10 & - & - & - & - & 3
   & 8 & 1 & 2 & 3 & 12 & - & 2 & - & 1 & 1 \\
\hline
U2 & 5 & 2 & 2 & 3 & 8 & - & 3 & 2 & - & -
   & 8 & 1 & 2 & 1 & 5 & - & 3 & 3 & - & 1 \\
\hline
U3 & 20 & 3 & - & 1 & 32 & - & 4 & 3 & - & -
   & 20 & 3 & 1 & - & 25 & - & 22 & 1 & - & 2 \\
\hline
U4 & 8 & 3 & 7 & 4 & 8 & 2 & 2 & - & 1 & 1
   & 7 & 9 & 6 & 4 & 21 & 4 & 8 & 2 & - & - \\
\hline
U5 & 12 & 2 & - & - & 10 & - & 6 & - & 1 & 5
   & 20 & - & 2 & - & 15 & - & 20 & - & 6 & 1 \\
\hline
U6 & 20 & - & - & - & 5 & - & 27 & - & 28 & -
   & 20 & 5 & 4 & 16 & - & - & 12 & - & 38 & 8 \\
\hline
U7 & 7 & 10 & 1 & 7 & 7 & - & - & - & 7 & 2
   & 6 & 1 & 4 & 3 & - & 16 & - & - & 6 & - \\
\hline
\textbf{Sum} 
& 76 & 22 & 10 & 17 & 80 & 2 & 42 & 5 & 37 & 11
& 89 & 20 & 21 & 27 & 78 & 20 & 67 & 6 & 51 & 13 \\
\hline
\end{tabular}
}
\label{tab:performance}
\end{table*}
\vspace{-25pt}
\paragraph{\textbf{RQ1: Performance Differences Between \textit{ArguMath} and Baseline}}
Overall, \textit{ArguMath} led to higher quantities and a broader diversity of theory-aligned interactions than the baseline system (Table~\ref{tab:performance}). Quantitative analysis of interaction logs showed that \textit{ArguMath} elicited more student responses (89 vs. 76), TRQF-coded teacher questions (68 vs. 49), and Toulmin-coded student responses (235 vs. 177). These results suggest that \textit{ArguMath} effectively promotes higher engagement and more frequent application of pedagogical frameworks compared to traditional training.

To assess diversity, we calculated normalized Shannon's entropy,   

$H = (-\sum_{i=1}^k p_i \ln p_i) / \ln k$,
yielding an evenness score between 0 and 1. \textit{ArguMath} achieved higher diversity scores than the baseline for both TRQF (0.99 vs. 0.96) and Toulmin (0.85 vs. 0.75) distributions. These results reflect more diverse pedagogical strategies and balanced argumentation. In particular, \textit{ArguMath} prompted more Teleological questions to elicit student methodologies, for instance, U2 asked, ``Why did you multiply by 0.3 instead of 30 or 3?''

\begin{figure}
    \centering
    \includegraphics[width=1\linewidth]{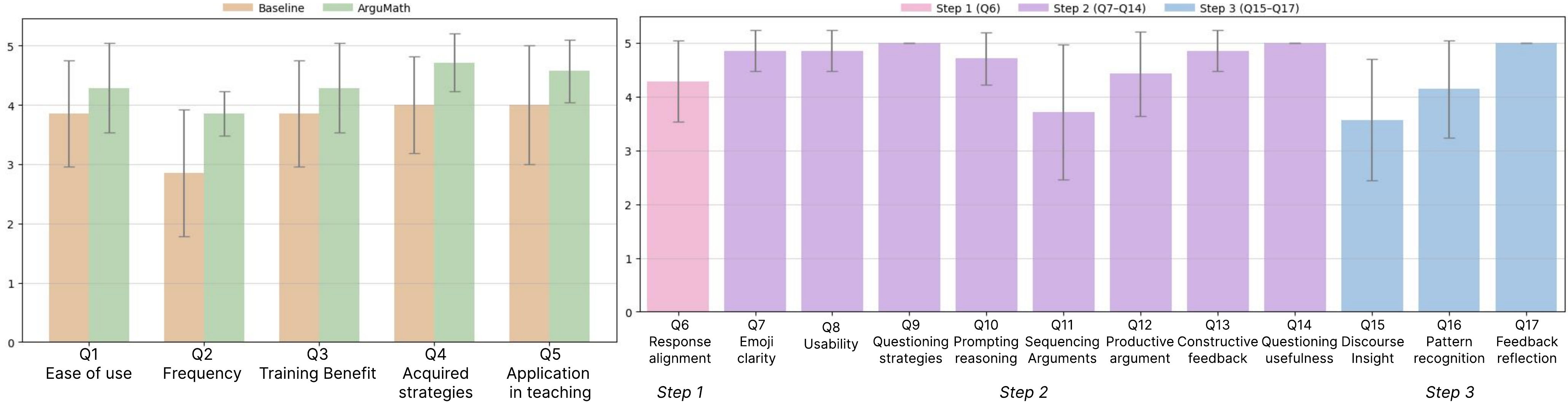}
    \caption{Means and standard errors ($5$-point Likert scale) for overall system comparison (Q1--Q5, left) and \textit{ArguMath}-specific feature ratings (Q6--Q17, right).}
    \label{fig:user_results}
\end{figure}

\paragraph{\textbf{RQ2: Perceptions of \textit{ArguMath} and Baseline}}
As shown in Figure~\ref{fig:user_results}, \textit{ArguMath} received higher ratings than the baseline across all measures, notably in future use ($M=3.86\,(0.38)$ vs.\ $M=2.86\,(1.07)$), argumentation orchestration practice ($M=4.29\, (0.76)$ vs. $M=3.86\, (0.90)$), perceived helpfulness for acquiring strategies ($M=4.71\, (0.49)$ vs. $M=4.00\, (0.82$)), and classroom applicability ($M=4.57$ vs. $4.00$).

Step-specific ratings confirm the efficacy of the classroom simulation (\textit{Step~2}) and strategy reflection (\textit{Step~3}). Participants highly rated \textit{ArguMath} for practicing questioning strategies (Q9: $M=5.00$), prompting students to articulate their reasoning (Q10: $M=4.71\, (0.49)$), and utilizing suggestions and feedback to guide instructional moves (Q12: $M=4.43\, (0.79)$; Q13: $M=4.86\, (0.38)$; Q14: $M=5.00\, (0.00)$). Qualitative feedback highlighted the system's realism; E1 noted that simulated responses mirrored actual student dialogue, such as ``oh, you mean like this.'' 
The interface was also perceived as intuitive (Q7--8: $M=4.86, (0.38)$), with U6 praising the clarity of visual elements.
However, lower ratings for sequencing student contributions (Q11: $M=3.71\, (1.25)$) suggest challenges in managing discourse flow. U1 reported feeling ``overwhelmed'' when multiple student answers appeared simultaneously.
Regarding \textit{Step~3}, features for noticing reasoning patterns (Q16: $M=4.14\, (0.90)$) and providing actionable feedback (Q17: $M=5.00\, (0.00)$) were well-received. E2 emphasized that labeling question types helped identify a lack of variety in her pedagogical approach.



%% file: sections/7-discussion.tex
\section{Discussion} 
Results suggested that \textit{ArguMath} supports the development of theory-aligned questioning strategies, and we further analyze the design implications as follows.


\textbf{\textit{Student Simulation using Authentic Classroom Discourse}} Realistic simulations of student struggles are crucial for effective pedagogical role-play. Participants rated the \textit{ArguMath} simulation as highly realistic, noting that varied engagement, informal utterances, and pauses closely mirrored actual classroom interactions. Such realism motivated participants to iteratively refine their questioning strategies. While simple prompt engineering often lacks student-like hesitation or reasoning~\cite{pan2025tutorup}, \textit{ArguMath} leverages Retrieval-Augmented Generation (RAG) based on 214 student profiles. This approach reflects diverse reasoning quality and participation patterns. Furthermore, unlike prior work limited to small-group chat interfaces~\cite{pan2025tutorup}, our emoji-based design supports classroom-scale argumentation. Participants valued the interface for its intuitiveness.

\textbf{\textit{Structured Theory-based Reflection}} A key feature of \textit{ArguMath} is its structured reflection scaffold, requiring PMTs to annotate their questions via TRQF and interpret student responses through Toulmin's model. Participants identified these features as the system's most valuable aspects (Figure~\ref{fig:user_results}). The structured scaffold shifted PMTs' participation from simply generating questions to active diagnosis of instructional moves. By examining how specific questioning types elicited argument components, PMTs engaged in a rigorous cycle of practice and reflection. Aligning with the ``approximations of practice'' framework~\cite{grossman2009teaching}, \textit{ArguMath} integrates teaching rehearsals with theoretically grounded analysis. However, effective implementation requires that PMTs are familiarized with the underlying pedagogical theories prior to the annotation activities.

\textbf{\textit{Limitation and Future Work}}
First, this study was conducted in a controlled setting with fixed grades and topics, potentially limiting generalizability. As simulated students were derived from a single dataset, they may not capture broader linguistic diversity. Future research should test \textit{ArguMath} across more diverse classroom dynamics. 
Second, the text-only interface restricts applicability to visually intensive domains like geometry or functions, where graphs and figures are critical for reasoning. Future iterations should incorporate equation and figure-support features.
Third, more sample sizes and longitudinal design are required in future studies to enable more robust statistical analyses.

%% file: sections/8-conclusion.tex
\section{Conclusion}
We present \textit{ArguMath}, an LLM-based classroom simulator that supports PMTs in orchestrating mathematical argumentation through context personalization, retrieval-augmented student simulation, and theory-based reflection and feedback. Findings from a within-subjects study and expert interviews suggest that these features help PMTs practice diverse theory-aligned questioning strategies, better understand students' mathematical reasoning, and strengthen PMTs' readiness to orchestrate mathematical argumentation. Together, the results indicate the potential of AI-simulated classrooms to support the integration of theory and practice in mathematical argumentation within teacher preparation.